\documentclass[12pt,english,draftcls, one column]{IEEEtran}
\usepackage[T1]{fontenc}
\usepackage[latin9]{inputenc}
\usepackage{float}
\usepackage{amsmath}
\usepackage{amsthm}
\usepackage{amssymb}
\usepackage{graphicx}
\usepackage{multirow}
\usepackage{xcolor}
\usepackage{subcaption}
\usepackage{multicol}
\usepackage{diagbox}
\usepackage{multirow}
\usepackage{cite}
\usepackage{algorithm, algorithmic}

\makeatletter

\floatstyle{ruled}
\newfloat{algorithm}{tbp}{loa}

\floatname{algorithm}{\protect Algorithm}

\theoremstyle{plain}

\theoremstyle{plain}

\ifCLASSINFOpdf
\else
\fi
\usepackage{babel}

\makeatother

\usepackage{babel}
\providecommand{\propositionname}{Proposition}
\providecommand{\theoremname}{Theorem}

\begin{document}


\title{Joint Precoding and Fronthaul Compression for Cell-Free MIMO Downlink With Radio Stripes}

\author{Sangwon Jo, \textit{Student Member}, \textit{IEEE}, Hoon Lee, \textit{Member}, \textit{IEEE}, \\and Seok-Hwan Park, \textit{Member}, \textit{IEEE}\vspace{-10mm}\thanks{


This work was supported by the National Research Foundation (NRF) of Korea funded by the MOE under Grant 2019R1A6A1A09031717, and MSIT under Grants 2021R1I1A3054575 and RS-2023-00238977, and in part by by Korea Research Institute for defense Technology planning and advancement (KRIT) grant
funded by the Korea government (DAPA(Defense Acquisition
Program Administration)) (21-106-A00-007, Space-Layer
Intelligent Communication Network Laboratory, 2022).

S. Jo and S.-H. Park are with the Division of Electronic Engineering, Jeonbuk
National University, Jeonju, Korea (email: tkddnjs9803@jbnu.ac.kr, seokhwan@jbnu.ac.kr).
H. Lee is with the Department of Information and Communications Engineering, Pukoyng National University, Busan, Korea (hoonlee@ieee.org).

Copyright (c) 2023 IEEE. Personal use of this material is permitted. However, permission to use this material for any other purposes must be obtained from the IEEE by sending a request to pubs-permissions@ieee.org.}}
\maketitle
\begin{abstract}
A sequential fronthaul network, referred to as radio stripes, is a promising fronthaul topology of cell-free MIMO systems. In this setup, a single cable suffices to connect access points (APs) to a central processor (CP). Thus, radio stripes are more effective than conventional star fronthaul topology which requires dedicated cables for each of APs. Most of works on radio stripes focused on the uplink communication or downlink energy transfer.
This work tackles the design of the downlink data transmission for the first time.
The CP sends compressed information of linearly precoded signals to the APs on fronthaul.
Due to the serial transfer on radio stripes, each AP has an access to all the compressed blocks which pass through it.
Thus, an advanced compression technique, called Wyner-Ziv (WZ) compression, can be applied in which each AP decompresses all the received blocks to exploit them for the reconstruction of its desired precoded signal as side information.
The problem of maximizing the sum-rate is tackled under the standard point-to-point (P2P) and WZ compression strategies. Numerical results validate the performance gains of the proposed scheme.
\end{abstract}
\theoremstyle{theorem}
\newtheorem{theorem}{Theorem}
\theoremstyle{proposition}
\newtheorem{proposition}{Proposition}
\theoremstyle{lemma}
\newtheorem{lemma}{Lemma}
\theoremstyle{corollary}
\newtheorem{corollary}{Corollary}
\theoremstyle{definition}
\newtheorem{definition}{Definition}
\theoremstyle{remark}
\newtheorem{remark}{Remark}


\section{Introduction}

Cell-free massive MIMO network is envisioned to mitigate the impact of inter-cell interference signals of cell boundary users. This can be made possible by moving the channel encoding and decoding functionality from distributed access points (APs) to a central processor (CP) \cite{Ngo:TWC17}. An essential requirement for cell-free MIMO systems is fronthaul networks which realize coordination among APs and CP through, for instance, optical fiber cables.
Many early works have considered a star fronthaul topology where each AP has a dedicated fronthaul connection to the CP \cite{Ngo:TWC17, Bjornson:TWC20}.
Albeit the success in the communication performance, the star fronthaul topology poses practical challenges as it requires a number of high-capacity long-length cables to realize individual fronthaul interconnection for all APs.

A bus fronthaul topology, also referred to as \textit{radio stripes}, has been regarded as a promising solution to implement practical cost-effective fronthaul systems \cite{Shaik-et-al:TC21, Chiotis:MeditCom22, Lopez:TWC22, Zhang:Entropy20}. In this setup, distributed APs are \textit{sequentially} connected to the CP with a single cable.
Compared to conventional star fronthaul topology, the radio stripe enables an easier deployment of cell-free MIMO systems, particularly for dense areas such as sports arenas and railway stations.

The viability of the radio stripe architecture has been investigated for uplink cell-free MIMO networks \cite{Shaik-et-al:TC21, Chiotis:MeditCom22}.
The design of linear processing at APs was addressed in \cite{Shaik-et-al:TC21}. The fronthaul links are simply assumed to deliver soft estimates of the uplink transmit signals and additional side information.
Compress-and-forward relaying strategies were adopted in  \cite{Chiotis:MeditCom22} to exploit finite-capacity radio stripes. The downlink cell-free MIMO systems with the radio stripes have also been studied \cite{Lopez:TWC22,Zhang:Entropy20}, but they have been confined to specific application scenarios such as energy transfer \cite{Lopez:TWC22} and a single receiving user equipment (UE) \cite{Zhang:Entropy20}. These scenarios have no adequate investigations on inter-UE interference management. Thus, the signal processing strategies of these conventional works cannot be straightforwardly extended to generic multi-UE downlink cell-free MIMO networks.

For the first time, this paper considers a generic downlink cell-free MIMO system with the radio stripes where separate APs, connected through the radio stripe fronthaul networks, provide donwlink communication services for multiple UEs.
Unlike the scenarios studied in \cite{Lopez:TWC22, Zhang:Entropy20}, which focused on energy transfer or single-UE setups, the design of fronthaul and wireless links should account for the impact of inter-UE interference signals.
To this end, the CP first performs a cooperative precoding across APs and sends the compressed information of precoded signals.
Since the compressed blocks are serially transferred on the radio stripes through APs,
each AP has a chance of decompressing all the blocks that pass through it.
To realize this novel fronthauling policy, we exploit the notion of the Wyner-Ziv (WZ) compression \cite{Xiong:SPM04, Park:SPM14}, in which each AP decompresses all the received compressed blocks so as to utilize them as side information for decompressing its transmit signal.
We tackle the sum-rate maximization problems with the standard point-to-point (P2P) compression \cite[Sec. 3.6]{Gamal:Cambridge11} and WZ compression techniques \cite{Xiong:SPM04, Park:SPM14} under the constraints on the per-AP transmit powers and fronthaul capacity.
To efficiently address this nonconvex problem, iterative algorithms based on the weighted minimum mean squared error (WMMSE) approach \cite{Christensen:TWC08} and a convexification method \cite{Zhou:TSP16} are derived.
The performance gains of the WZ compression scheme compared to the P2P compression are investigated via numerical results.

\section{System Model\label{sec:System-Model}}

\begin{figure}
\centering\includegraphics[width=0.6\linewidth]{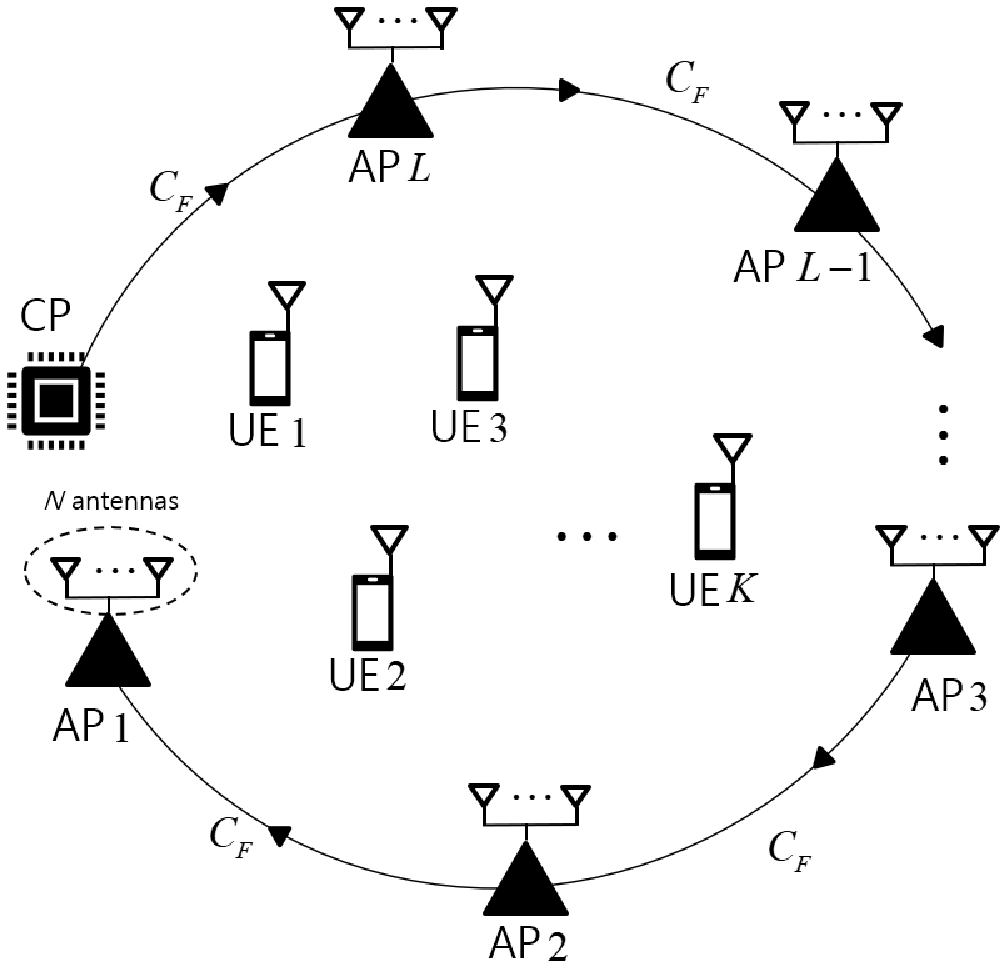}
\vspace{-3mm}
\caption{\label{fig:system-model}Illustration of the downlink of cell-free MIMO system with a radio stripe fronthaul network. A single fronthaul cable connects the CP to AP $L$, AP $L-1$, $\ldots$, AP 2, and AP 1.}
\end{figure}

We consider the downlink of a cell-free MIMO system in which
$K$ single-antenna UEs are served by a CP through $L$ APs each equipped with $N$ antennas.
Let $\mathcal{K} = \{1,2,\ldots, K\}$ and $\mathcal{L} = \{1,2,\ldots, L\}$ be sets of UE and AP indices, respectively.
As shown in Fig. \ref{fig:system-model}, $L$ APs and the CP are connected by the radio stripes \cite{Shaik-et-al:TC21} so that the fronthaul connection goes from the CP to AP $L$, AP $L-1$, ..., to AP $1$. For simplicity, we denote the CP as AP $L+1$. Let $\mathtt{F}_{i}$ be the fronthaul link from AP $i$ to AP $i-1$ by $\mathtt{F}_i$ for
$i\in\{2,3,\ldots,L,L+1\}$. Then, $\mathtt{F}_{L+1}$ represents the fronthaul link from the CP to AP $L$.
Every fronthaul link $\mathtt{F}_i$ has a finite capacity $C_F$ in bits/symbol, i.e., the number of bits delivered per symbol period.

Let $\mathbf{h}_{k,i}\sim\mathcal{CN}(\mathbf{0},\rho_{k,i}\mathbf{I})\in\mathbb{C}^{N}$ be the Rayleigh fading channel vector from AP $i$ to UE $k$ with the path-loss $\rho_{k,i}$. Then, received signal $y_k\in\mathbb{C}$ of UE $k$ is given as
\begin{align}
    y_k = \mathbf{h}_k^H \mathbf{x} + z_k, \label{eq:received-signal-downlink}
\end{align}
where $\mathbf{h}_k = [\mathbf{h}_{k,1}^H \, \cdots \,\mathbf{h}_{k,L}^H]^H\in\mathbb{C}^{NL\times 1}$ denotes the channel vector from all APs to UE $k$, $\mathbf{x} = [\mathbf{x}_1^H \, \cdots \, \mathbf{x}_L^H]^H \in \mathbb{C}^{NL\times 1}$ stacks the transmit signal vectors all APs with $\mathbf{x}_i\in\mathbb{C}^{N\times 1}$ being the signal transmitted by AP $i$, and $z_k\sim\mathcal{CN}(0, \sigma_z^2)$ represents the additive noise.
The transmit signal of each AP $i$ is subject to a power constraint
\begin{align}
    \mathbb{E}\left[ \|\mathbf{x}_i\|^2 \right] \leq P_{\text{tx}}. \label{eq:power-constraint}
\end{align}
We define the transmit signal-to-noise ratio (SNR) of the downlink channel as $P_{\text{tx}}/\sigma_z^2$

The CP designs the signal processing strategy of all APs based on the estimated channel state information (CSI). Let $\hat{\mathbf{h}}_{k,i}$ be an estimate of $\mathbf{h}_{k,i}$, which can be written as
\begin{align}
    \mathbf{h}_{k,i} = \hat{\mathbf{h}}_{k,i} + \mathbf{e}_{k,i}, \label{eq:additive-CSI-error}
\end{align}
with the CSI error vector $\mathbf{e}_{k,i}$.
With the linear minimum mean squared error estimator, $\mathbf{e}_{k,i}$ becomes uncorrelated with $\hat{\mathbf{h}}_{k,i}$ and distributed as $\mathbf{e}_{k,i}\sim\mathcal{CN}(\mathbf{0}, \beta_{k,i}\mathbf{I})$ \cite{Choi:TWC20}.
The error variance $\beta_{k,i} \in [0, \rho_{k,i}]$ increases as the pilot length gets shorter or the SNR of the uplink training channel decreases.
Let us define the nominal channel vector $\hat{\mathbf{h}}_k = [\hat{\mathbf{h}}_{k,1}^H \hat{\mathbf{h}}_{k,2}^H \cdots \hat{\mathbf{h}}_{k,L}^H]^H$ and CSI error vector $\mathbf{e}_k=[\mathbf{e}_{k,1}^H \mathbf{e}_{k,2}^H \cdots \mathbf{e}_{k,L}^H]^H$ for UE $k$.
Then, the error vector $\mathbf{e}_k$ is distributed as $\mathbf{e}_k \sim \mathcal{CN}(\mathbf{0}, \mathbf{E}_k)$ with $\mathbf{E}_k = \text{blkdiag}(\{\beta_{k,i}\mathbf{I}\}_{i\in\mathcal{L}})$.

The downlink transmission from the CP to the UEs consists of two phases: \textit{i)} CP-to-APs fronthauling phase; and \textit{ii)} APs-to-UEs wireless accessing phase.
In the fronthauling phase, the CP sends bit streams, which represent quantized versions of the precoded baseband signals, to the APs via the radio stripes.
In the wireless accessing phase, the APs simultaneously transmit the downlink signals over wireless channels toward UEs, where the downlink signals are obtained by decompressing the bit streams received through the fronthaul links.
We denote the time durations of the fronthauling and wireless accessing transmission phases by $n_F$ and $n_A$, respectively, which are measured in the symbol period.
Thus, $n_A$ becomes the number of channel uses for downlink transmission which
corresponds to the blocklength of the channel coding.
The ratio between two phases is defined as
$r = n_F / n_A$.

\section{Precoding and Fronthaul Compression With Radio Stripes}

This section proposes a sequential fronthauling strategy to compute efficient precoding solutions through the radio stripes. Let $s_{k}\sim\mathcal{CN}(0,1)$ be the data symbol intended to UE $k$. To obtain the transmit signal vector $\mathbf{x}$ in \eqref{eq:received-signal-downlink}, the CP first generates an intermediate vector $\tilde{\mathbf{x}}\in\mathbb{C}^{NL\times1}$ given by
\begin{align}
    \tilde{\mathbf{x}} = \sum\nolimits_{k\in\mathcal{K}} \mathbf{v}_k s_k \triangleq [\tilde{\mathbf{x}}_1^H\ \cdots\ \tilde{\mathbf{x}}_L^H]^H, \label{eq:precoding}
\end{align}
where $\mathbf{v}_k\in\mathbb{C}^{NL\times 1}$ is the linear precoding vector for UE $k$ and the $i$-th subvector $\tilde{\mathbf{x}}_{i}\in\mathbb{C}^{N\times1}$ of $\tilde{\mathbf{x}}$ indicates the transmit signal at AP $i$. For the sequential fronthauling, the CP delivers the signal vector $\tilde{\mathbf{x}}_{i}$ to AP $i$ through APs $L,L-1,\cdots,i+1$.
Since the fronthaul links have finite capacity $C_F$ bits/symbol, each precoded signal $\tilde{\mathbf{x}}_i$ is quantized and compressed.
Thus, the signal vector $\mathbf{x}_i$ transmitted by AP $i$ is a quantized version of $\hat{\mathbf{x}}_i$ and is modeled as
\begin{align}
    \mathbf{x}_i = \tilde{\mathbf{x}}_i + \mathbf{q}_i. \label{eq:quantization}
\end{align}
Under the Gaussian test channel \cite{Zhou:TSP16}, the quantization noise $\mathbf{q}_i$ is independent of $\tilde{\mathbf{x}}_i$ and distributed as $\mathbf{q}_i\sim\mathcal{CN}(\mathbf{0}, \boldsymbol{\Omega}_i)$.
The power constraint (\ref{eq:power-constraint}) is rewritten as
\begin{align}
    \sum\nolimits_{k\in\mathcal{K}} \left\|\mathbf{D}_i^H\mathbf{v}_k\right\|^2 + \text{tr}\left(\boldsymbol{\Omega}_i\right) \leq P_{\text{tx}}, \label{eq:power-constraint-rewritten}
\end{align}
where $\mathbf{D}_i\in\mathbb{C}^{NL\times N}$ is filled with zeros except for the rows from $N(i-1)+1$ to $Ni$ being an identity matrix of size $N$.

\begin{table}
\centering%
\begin{tabular}{|c||c|c|c|c|}
\hline
 & $\!\!\!\!\!\!\begin{array}{c}\mathtt{F}_{L+1} \\ \text{\scriptsize (CP $\to$ AP $L$)} \end{array}\!\!\!\!\!\!$ & $\!\!\!\!\!\!\begin{array}{c}\mathtt{F}_{L} \\ \text{\scriptsize (AP $L$ $\to$ AP $L-1$)} \end{array}\!\!\!\!\!\!$ & $\cdots$ & $\!\!\!\!\!\!\begin{array}{c}\mathtt{F}_{1} \\ \text{\scriptsize (AP $2$ $\to$ AP $1$)} \end{array}\!\!\!\!\!\!$ \tabularnewline
\hline
\hline
Slot 1 & $\mathtt{B}_1$ & -- & $\cdots$ & -- \tabularnewline
\hline
Slot 2 & $\mathtt{B}_2$ & $\mathtt{B}_1$ & $\ddots$ & $\vdots$\tabularnewline
\hline
$\vdots$ & $\vdots$ & $\vdots$ & $\ddots$ & -- \tabularnewline
\hline
Slot $L$ & $\mathtt{B}_L$ & $\mathtt{B}_{L-1}$ & $\cdots$ &  $\mathtt{B}_1$\tabularnewline
\hline
\end{tabular}
~

~

~
Table I: Compressed blocks $\mathtt{B}_1,\ldots,\mathtt{B}_L$ transferred on the fronthaul links $\mathtt{F}_1,\ldots,\mathtt{F}_L$ at time slots $t=1,2,\ldots,L$
\end{table}

As illustrated in Table I, the time-division multiple access protocol is adopted for the fronthauling phase.
We define $\mathtt{B}_i$ as the compressed block of the quantized signal $\mathbf{x}_i$, which needs to be delivered to AP $i$.
The time duration of the fronthauling phase is divided into $L$ time slots, where each time slot spans $n_F/L$ symbol periods.
In time slot 1, the fronthaul link $\mathtt{F}_{L+1}$ connecting the CP to AP $L$ carries the block $\mathtt{B}_1$ which is intended for AP $1$.
Subsequently, in time slot 2, $\mathtt{B}_1$ is transferred from AP $L$ to AP $L-1$ on the fronthaul $\mathtt{F}_L$. At the same time, AP $L-1$ receives $\mathtt{B}_2$ from the CP on $\mathtt{F}_{L+1}$.
In general, in time slot $t$ ($t=1,2,\ldots,L$), the block $\mathtt{B}_i$ with $i\leq t$ moves on the fronthaul link $\mathtt{F}_{L+i-t+1}$ from AP $L+i-t+1$ to AP $L+i-t$.

Once the fronthaul transmission over the $L$ slots is completed, each AP $i$ has an access to its coded block $\mathtt{B}_i$.
Thus, it can decompress $\mathtt{B}_i$ to obtain the transmit signal vector $\mathbf{x}_i$, which is a quantized version of $\tilde{\mathbf{x}}_i$ as in (\ref{eq:quantization}), and transmits it over the downlink wireless channel. Thus, the received signal $y_{k}$ in \eqref{eq:received-signal-downlink} can be expressed as
\begin{align}    y_{k}&=\hat{\mathbf{h}}_{k}^{H}\mathbf{v}_{k}s_{k} + \sum\nolimits_{l\in\mathcal{K}\setminus\{k\}}\hat{\mathbf{h}}_{k}^{H}\mathbf{v}_{l}s_{l} +\sum\nolimits_{l\in\mathcal{K}}\mathbf{e}_{k}^{H}\mathbf{v}_{l}s_{l} +\mathbf{h}_{k}^{H}\mathbf{q}+z_{k},
    \label{eq:yk}
\end{align}
where $\mathbf{q}=[\mathbf{q}_{1}^{H}\cdots\mathbf{q}_{L}^{H}]^{H}$ stands for the quantization noise in the fronthauling phase distributed as $\mathbf{q}\sim\mathcal{CN}(\mathbf{0}, \bar{\boldsymbol{\Omega}})$ with $\bar{\boldsymbol{\Omega}} = \text{blkdiag}(\{\boldsymbol{\Omega}_i\}_{i\in\mathcal{L}})$. In \eqref{eq:yk}, the second and third terms stand for the inter-UE interference and the noise induced by the CSI estimation error $\mathbf{e}_{k}$, respectively. The achievable data rate of UE $k$ can be written as \cite{Choi:TWC20}
\begin{align}
    R_k = \log_2\left( 1 + \frac{|\hat{\mathbf{h}}_k^H\mathbf{v}_k |^2}{ \mathtt{IN}_k\left(\mathbf{v}, \boldsymbol{\Omega}\right)  } \right), \label{eq:data-rate-UE-k}
\end{align}
where the interference-plus-noise power, denoted by $\mathtt{IN}_k(\mathbf{v}, \boldsymbol{\Omega})$, is defined as
\begin{align}
    \mathtt{IN}_k\left(\mathbf{v}, \boldsymbol{\Omega}\right) & = \sum\nolimits_{l\in\mathcal{K}\setminus\{k\}} \left|\hat{\mathbf{h}}_k^H\mathbf{v}_l\right|^2 + \sum\nolimits_{l\in\mathcal{K}} \mathbf{v}_l^H\mathbf{E}_k\mathbf{v}_l \nonumber \\
    & + \text{tr}\left(\left(\hat{\mathbf{h}}_k\hat{\mathbf{h}}_k^H + \mathbf{E}_k \right)\bar{\boldsymbol{\Omega}
    }\right) + \sigma_z^2 \label{eq:IF-plus-noise-power}
\end{align}
with $\boldsymbol{\Omega}=\{\boldsymbol{\Omega}_{k}\}_{i\in\mathcal{L}}$.

\section{Fronthaul Compression Strategies and Problem Definition} \label{sec:strategy-problem}

In this section, we describe two fronthaul compression strategies: \textit{i)} \textit{Wyner-Ziv} (WZ) compression \cite{Xiong:SPM04}; and \textit{ii)} \textit{point-to-point} (P2P) compression \cite[Sec. 3.6]{Gamal:Cambridge11}. It is then followed by the description of the joint optimization task for fronthauling and accessing phases.

\subsection{WZ Compression} \label{sub:distributed-compression}

Due to the nature of the sequential fronthauling phase, AP $i$ receives the preceding blocks $\mathtt{B}_1,\mathtt{B}_2,\ldots,\mathtt{B}_{i-1}$ before receiving its desired block $\mathtt{B}_i$. Consequently, AP $i$ has a chance of recovering the signals $\mathbf{x}_1,\mathbf{x}_2,\ldots,\mathbf{x}_{i-1}$. As depicted in Fig. \ref{fig:decompression-operation}(a), these preceding signals can be exploited as side information for the decompression of $\mathtt{B}_i$
The WZ compression, also called distributed compression, can be applied to the design of compressor and decompressor for this scenario which enables the compressor at the CP to use a finer quantizer \cite{Xiong:SPM04, Park:SPM14}.
The number of bits, i.e., the minimal size of block $\mathtt{B}_i$, needed for successful decompression with the WZ compression scheme is given as $n_A \cdot g_{\text{WZ},i}(\mathbf{v}, \boldsymbol{\Omega})$ bits, where $g_{\text{WZ},i}(\mathbf{v}, \boldsymbol{\Omega})$ measures the compression rate in the number of bits per sample. The function $g_{\text{WZ},i}(\mathbf{v}, \boldsymbol{\Omega})$ is defined as
\begin{align}
    g_{\text{WZ},i}\left(\mathbf{v}, \boldsymbol{\Omega}\right) &= I\left( \tilde{\mathbf{x}}_i ; \mathbf{x}_i \big|  \mathbf{x}_1,\mathbf{x}_2,\ldots,\mathbf{x}_{i-1} \right) \nonumber \\
    &\overset{(\text{a)}}{=} \log_2\det\left( \bar{\mathbf{D}}_i^H \mathbf{V}_{\Sigma} \bar{\mathbf{D}}_i + \bar{\boldsymbol{\Omega}}_i \right) - \log_2\det\!\left( \bar{\mathbf{D}}_{i-1}^H \mathbf{V}_{\Sigma} \bar{\mathbf{D}}_{i-1} + \bar{\boldsymbol{\Omega}}_{i-1} \right)  \nonumber \\
    &  - \log_2\!\det\!\left(\boldsymbol{\Omega}_i\right), \label{eq:compression-rate-WZ}
\end{align}
where $\mathbf{V}_{\Sigma} \triangleq \sum_{l\in\mathcal{K}}\mathbf{v}_l\mathbf{v}_l^H$. Here, we have defined the matrices $\bar{\mathbf{D}}_i = [\mathbf{D}_1 \,\mathbf{D}_2\,\cdots\, \mathbf{D}_i] \in\mathbb{C}^{NL\times Ni}$ and $\bar{\boldsymbol{\Omega}}_i = \text{blkdiag}(\boldsymbol{\Omega}_1,\boldsymbol{\Omega}_2,\ldots,\boldsymbol{\Omega}_i)\in\mathbb{C}^{Ni\times Ni}$.
The equality (a) can be shown by using the chain rule of mutual information.

The block $\mathtt{B}_i$ of length $n_A \cdot g_{\text{WZ}, i}(\mathbf{v},\boldsymbol{\Omega})$ bits is delivered via each fronthaul link of capacity $C_F$ bit/symbol over $n_F/L$ symbol periods. Thus, for successful decompression of $\mathtt{B}_i$ with the WZ compression, the beamforming vector $\mathbf{v}=\{\mathbf{v}_{k}\}_{k\in\mathcal{K}}$ and the quantization noise covariance matrix $\boldsymbol{\Omega}$ should satisfy the following fronthaul capacity constraint:
\begin{align}
    g_{\text{WZ},i}(\mathbf{v}, \boldsymbol{\Omega})\leq\frac{r C_{F}}{L}. \label{eq:fronthaul-constraint-WZ}
\end{align}

\subsection{P2P Compression} \label{sub:P2P-compression}

\begin{figure}
        \centering
            \centering
            \includegraphics[width=0.9\linewidth]{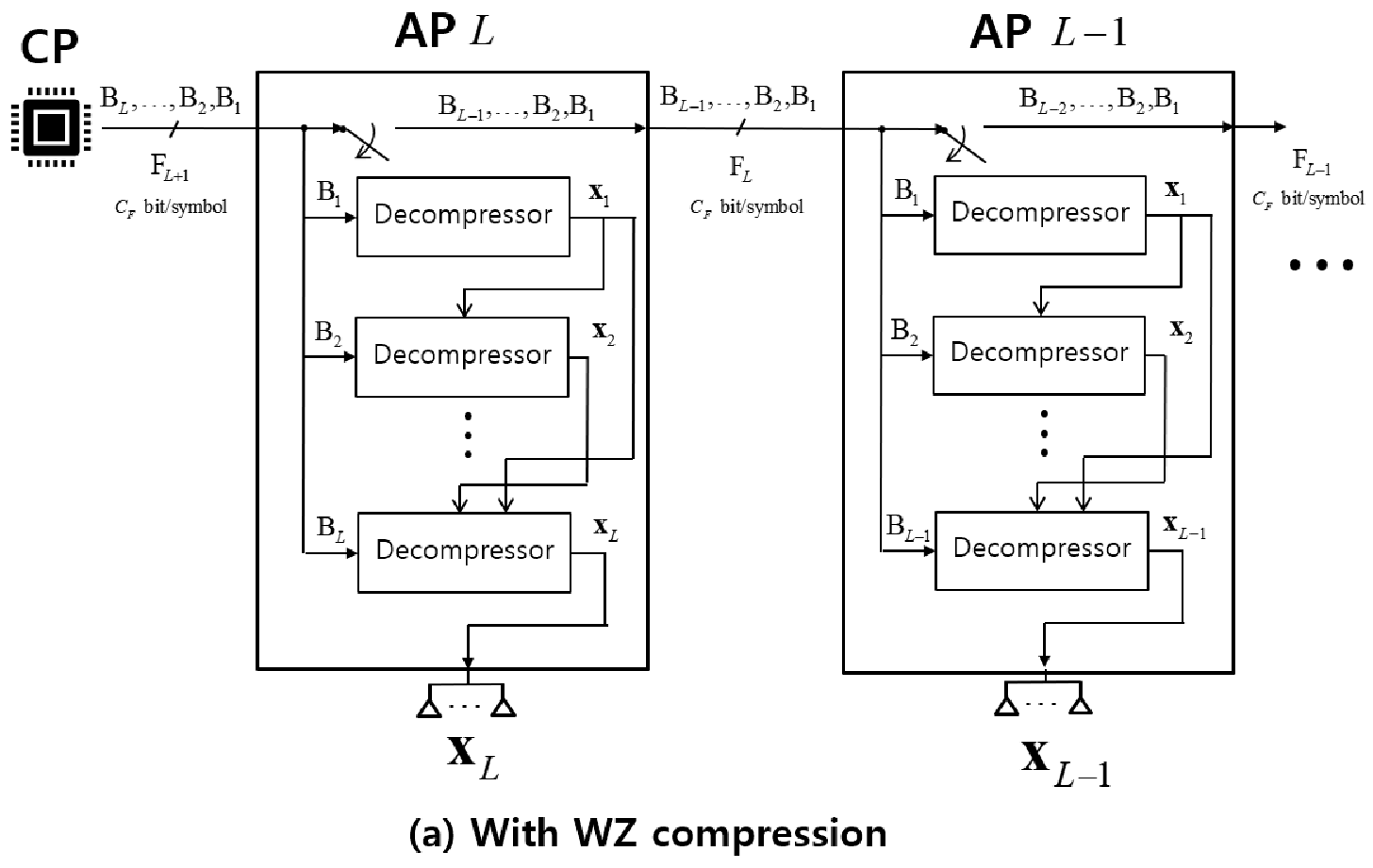}
        \hfill
        \\
        \vspace{3mm}
            \centering
            \includegraphics[width=0.9\linewidth]{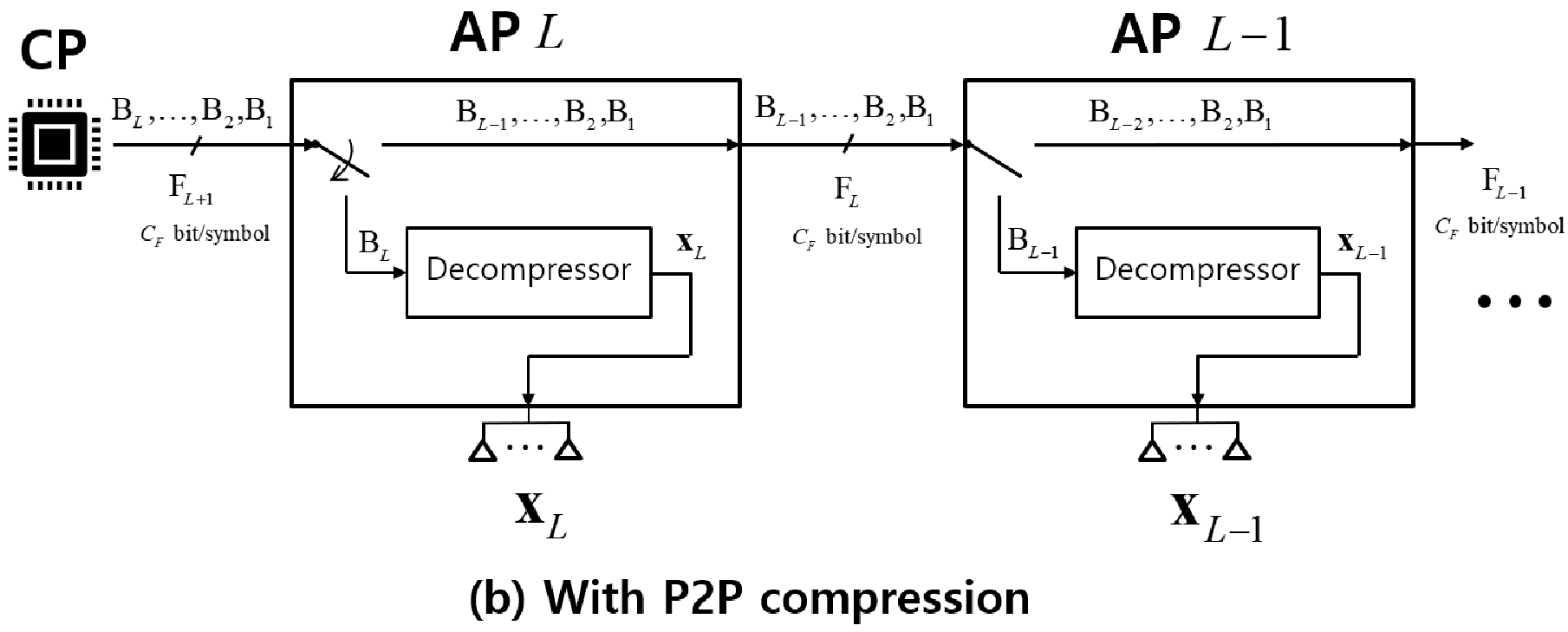}
        \caption{Comparison of the decompression operations at the APs with WZ and P2P compression schemes}
        \label{fig:decompression-operation}
\end{figure}

As shown in Fig. \ref{fig:decompression-operation}(b), in the P2P compression scheme, AP $i$ only decodes its desired block $\mathtt{B}_{i}$ while simply bypassing others $\mathtt{B}_{j}$, $\forall j<i$.
Therefore, the P2P compression strategy, which does not exploit the side information about the preceding APs, can be viewed as a special case of the WZ approach. Likewise \eqref{eq:fronthaul-constraint-WZ}, in the P2P compression scheme, the fronthaul capacity constraint can be expressed as
\begin{align}
    g_{\text{P2P}, i}\left(\mathbf{v},\boldsymbol{\Omega}\right) \leq \frac{ r C_F }{L}. \label{eq:fronthaul-constraint-P2P}
\end{align}
where the compression rate function for the P2P compression scheme $g_{\text{P2P},i}(\mathbf{v}, \boldsymbol{\Omega})$ is given by
\begin{align}
    g_{\text{P2P},i}\left(\mathbf{v}, \boldsymbol{\Omega}\right) & = I\left( \tilde{\mathbf{x}}_i ; \mathbf{x}_i \right) \nonumber \\
    &= \log_2\det\left( \mathbf{D}_i^H\mathbf{V}_{\Sigma}\mathbf{D}_i + \boldsymbol{\Omega}_i \right) - \log_2\det \left(\boldsymbol{\Omega}_i\right). \label{eq:compression-rate-P2P}
\end{align}

\subsection{Problem Description}

To optimize the fronthauling and accessing phases jointly, we tackle the problem of optimizing the precoding vector $\mathbf{v}$ and the fronthaul quantization policy $\boldsymbol{\Omega}$, with the goal of maximizing the sum-rate $\sum_{k\in\mathcal{K}} R_k$. The sum-rate maximization problem is stated as
\begin{subequations} \label{eq:problem-P2P}
\begin{align}
    \underset{\mathbf{v}, \boldsymbol{\Omega}, \mathbf{R}} {\mathrm{max.}}\,\,\, & \sum\nolimits_{k\in\mathcal{K}} R_k \, \label{eq:problem-P2P-cost} \\
 \mathrm{s.t. }\,\,\,\,\,\, & R_k \leq \log_2\left( 1 + \frac{|\hat{\mathbf{h}}_k^H\mathbf{v}_k |^2}{ \mathtt{IN}_k\left(\mathbf{v}, \boldsymbol{\Omega}\right)  } \right), \, k\in\mathcal{K}, \label{eq:problem-P2P-rate} \\
 & g_{X,i}\left(\mathbf{v},\boldsymbol{\Omega}\right) \leq \frac{ r C_F }{L},\,  i\in\mathcal{L}, \label{eq:problem-P2P-fronthaul} \\
 & \sum\nolimits_{k\in\mathcal{K}} \left\|\mathbf{D}_i^H\mathbf{v}_k\right\|^2 + \text{tr}\left(\boldsymbol{\Omega}_i\right) \leq P_{\text{tx}}, \, i\in\mathcal{L}, \label{eq:problem-P2P-power}
\end{align}
\end{subequations}
where $\mathbf{R}=\{\mathbf{R}_{k}\}_{k\in\mathcal{K}}$ denotes the achievable rate vector and left-hand-side $g_{X,i}\left(\mathbf{v},\boldsymbol{\Omega}\right)$ of the fronthaul capacity constraint in \eqref{eq:problem-P2P-rate} can be chosen as $X\in\{\text{WZ},\text{P2P}\}$, i.e., it is set to either \eqref{eq:fronthaul-constraint-WZ} or \eqref{eq:fronthaul-constraint-P2P}. Due to the constraints in \eqref{eq:problem-P2P-rate} and \eqref{eq:problem-P2P-fronthaul}, the above problem becomes nonconvex in general, and thus it is difficult to obtain the globally optimal solution.
It is not difficult to see that $g_{\text{WZ},i}(\mathbf{v},\boldsymbol{\Omega}) \leq g_{\text{P2P},i}(\mathbf{v},\boldsymbol{\Omega})$, and thus (\ref{eq:fronthaul-constraint-P2P}) is stricter than (\ref{eq:fronthaul-constraint-WZ}). Consequently, the feasible set of problem \eqref{eq:problem-P2P} for the WZ scheme always contains that of the P2P compression, implying that the WZ compression is superior to the P2P scheme.

\section{Optimization Algorithms} \label{sec:iterative-algorithm}

This section presents optimization algorithms to solve problem (\ref{eq:problem-P2P}) for the WZ and P2P compression schemes.

\subsection{WZ Compression}
We first consider the WZ compression where constraint \eqref{eq:problem-P2P-fronthaul} boils down to \eqref{eq:fronthaul-constraint-WZ}.
The problem (\ref{eq:problem-P2P}) is nonconvex due to the constraints (\ref{eq:problem-P2P-rate}) and (\ref{eq:problem-P2P-fronthaul}).
We handle the nonconvex access link rate constraint (\ref{eq:problem-P2P-rate}) with the WMMSE approach (see, e.g., \cite{Christensen:TWC08}).
The constraint (\ref{eq:problem-P2P-rate}) is satisfied if there exist $u_k\in\mathbb{C}$ and $w_k > 0$ that satisfy
\begin{align}
    R_k \leq \log_2 w_k - \frac{1}{\ln 2} w_k \, e_k( \mathbf{v}, \boldsymbol{\Omega}, u_k ) + \frac{1}{\ln 2}, \, k\in\mathcal{K}, \label{eq:inequality-WMMSE-1}
\end{align}
where $u_{k}$ and $w_{k}$ indicate scalar receiver filter and weight for UE $k$, respectively, and the mean squared error (MSE) function $e_k(\mathbf{v}, \boldsymbol{\Omega}, u_k)$ is defined as
\begin{align}
    e_k(\mathbf{v}, \boldsymbol{\Omega}, u_k) &= \mathbb{E}\left[ \big| s_k - u_k^H y_k \big|^2 \right] \nonumber \\
    &= \big| 1 - u_k^H\hat{\mathbf{h}}_k^H \mathbf{v}_k \big|^2 + |u_k|^2 \mathtt{IN}_k\left(\mathbf{v},\boldsymbol{\Omega}\right). \label{eq:MSE-k}
\end{align}
The condition (\ref{eq:inequality-WMMSE-1}) becomes equivalent to the original rate constraint (\ref{eq:problem-P2P-rate}) when the receive filter $u_k$ and the weight $w_k$ are respectively set to
\begin{align}
    u_k = \frac{ \hat{\mathbf{h}}_k^H\mathbf{v}_k }{ \big|\hat{\mathbf{h}}_k^H \mathbf{v}_k\big|^2 + \mathtt{IN}_k(\mathbf{v}, \boldsymbol{\Omega}) } \text{ and }
    w_k = \frac{1}{ e_k(\mathbf{v}, \boldsymbol{\Omega}, u_k) }. \label{eq:optimal-w-u}
\end{align}

The nonconvex constraint in (\ref{eq:problem-P2P-fronthaul}) with $X=\text{WZ}$ imposes that (\ref{eq:fronthaul-constraint-WZ}) is satisfied for all $i\in\mathcal{L}$.
By substituting the constraint (\ref{eq:fronthaul-constraint-WZ}) with $i=j$ into that with $i=j+1$ for $j=1,2,\ldots,L-1$ sequentially, it becomes
\begin{align}
    \log_2\det\left( \bar{\mathbf{D}}_i^H \mathbf{V}_{\Sigma} \bar{\mathbf{D}}_i + \bar{\boldsymbol{\Omega}}_i \right) - \sum\nolimits_{j=1}^i \log_2\det\left( \boldsymbol{\Omega}_j \right) \leq \frac{i r C_F}{L}, \, i\in\mathcal{L}. \label{eq:fronthaul-constraint-WZ-trick}
\end{align}
It has been reported in \cite{Zhou:TSP16} that (\ref{eq:fronthaul-constraint-WZ-trick}) holds if there exist a group of positive definite matrices $\boldsymbol{\Theta}_i \in \mathbb{C}^{Ni\times Ni}$, $i\in\mathcal{L}$, that satisfy
\begin{align}
    &\log_2\det\left( \boldsymbol{\Theta}_i \right) + \frac{1}{\ln 2} \text{tr}\left( \boldsymbol{\Theta}_i^{-1} \left( \bar{\mathbf{D}}_i^H \mathbf{V}_{\Sigma} \bar{\mathbf{D}}_i + \bar{\boldsymbol{\Omega}}_i \right) \right) - \frac{Ni}{\ln 2} \nonumber \\
    &\,\,\,\,\,\,- \sum\nolimits_{j=1}^i\log_2\det\left( \boldsymbol{\Omega}_{j} \right) \leq \frac{irC_F}{L}, \, i\in\mathcal{L}. \label{eq:convexified-fronthaul-constraint-WZ}
\end{align}
We can make the constraint (\ref{eq:convexified-fronthaul-constraint-WZ}) equivalent to the original one (\ref{eq:fronthaul-constraint-WZ}) by setting $\boldsymbol{\Theta}_i$ to
\begin{align}
    \boldsymbol{\Theta}_i = \bar{\mathbf{D}}_i^H \mathbf{V}_{\Sigma} \bar{\mathbf{D}}_i + \bar{\boldsymbol{\Omega}}_i. \label{eq:optimal-Theta-WZ}
\end{align}
Thus, for the WZ compression scheme, the sum-rate maximization task \eqref{eq:problem-P2P} can be equivalently transformed into
\begin{align}
    \underset{\mathbf{v}, \boldsymbol{\Omega}, \mathbf{R}, \mathbf{u}, \mathbf{w}, \boldsymbol{\Theta}} {\mathrm{max.}}\,\,\, & \sum\nolimits_{k\in\mathcal{K}} R_k \, \label{eq:problem-WZ-modified}\\
 \mathrm{s.t. }\,\,\,\,\,\, & R_k \ln 2 \leq \ln w_k - w_k \, e_k( \mathbf{v}, \boldsymbol{\Omega}, u_k ) + 1, \, k\in\mathcal{K}, \nonumber\\
 & \ln\det\left( \boldsymbol{\Theta}_i \right) + \text{tr}\left( \boldsymbol{\Theta}_i^{-1} \left( \bar{\mathbf{D}}_i^H \mathbf{V}_{\Sigma} \bar{\mathbf{D}}_i + \bar{\boldsymbol{\Omega}}_i \right) \right) - Ni \nonumber \\
 &\,\,\,\,\,\,- \sum\nolimits_{j=1}^i\ln\det\left( \boldsymbol{\Omega}_{j} \right) \leq \frac{irC_F}{L}\ln 2,\,  i\in\mathcal{L}, \nonumber\\
 & \sum\nolimits_{k\in\mathcal{K}} \left\|\mathbf{D}_i^H\mathbf{v}_k\right\|^2 + \text{tr}\left(\boldsymbol{\Omega}_i\right) \leq P_{\text{tx}}, \, i\in\mathcal{L}, \nonumber
\end{align}
where we have defined $\mathbf{u}=\{u_{k}\}_{k\in\mathcal{K}}$, $\mathbf{w}=\{w_{k}\}_{k\in\mathcal{K}}$, and $\boldsymbol{\Theta}=\{\boldsymbol{\Theta}_{i}\}_{i\in\mathcal{L}}$.

We summarize an optimization process for solving \eqref{eq:problem-WZ-modified} in Algorithm~1. Note that if the auxiliary variables $\mathbf{u}$, $\mathbf{w}$, and $\boldsymbol{\Theta}$ are fixed in the problem (\ref{eq:problem-WZ-modified}), the problem becomes a convex problem whose solution can be efficiently found with standard convex solvers such as, e.g., the CVX software \cite{Grant:CVX20}.
Also, the optimal $\mathbf{u}$, $\mathbf{w}$, and $\boldsymbol{\Theta}$ for fixed primary variables $\mathbf{v}$ and $\boldsymbol{\Omega}$ are given in closed-forms as (\ref{eq:optimal-w-u}) and (\ref{eq:optimal-Theta-WZ}).
Therefore, we can monotonically increase the objective sum-rate $\sum_{k\in\mathcal{K}} R_k$ by alternately optimizing one of the variable sets $\{\mathbf{v},\boldsymbol{\Omega}\}$ and $\{\mathbf{u},\mathbf{w},\boldsymbol{\Theta}\}$ for fixed other. Such an alternating optimization process is repeated until the convergence.

\begin{algorithm}
\caption{Alternating optimization algorithm for problem \eqref{eq:problem-P2P} under the WZ compression}

\textbf{\footnotesize{}1}\textbf{ Initialize:} Set $t\leftarrow1$, and initialize $\{\mathbf{v},\boldsymbol{\Omega}\}$ to an arbitrary point that satisfies (\ref{eq:problem-P2P-rate})-(\ref{eq:problem-P2P-power});

\textbf{\footnotesize{}2}\textbf{ repeat}

\textbf{\footnotesize{}3}~~~~$t\leftarrow t+1$;

\textbf{\footnotesize{}4}~~~~update $\{\mathbf{u},\mathbf{w},\boldsymbol{\Theta}\}$ according to (\ref{eq:optimal-w-u}) and (\ref{eq:optimal-Theta-WZ}) with $\{\mathbf{v},\boldsymbol{\Theta}\}$ being fixed;

\textbf{\footnotesize{}5}~~~update $\{\mathbf{v},\boldsymbol{\Omega}\}$
as a solution of the problem (\ref{eq:problem-WZ-modified}) with $\{\mathbf{u},\mathbf{w},\boldsymbol{\Theta}\}$ being fixed;

\textbf{\footnotesize{}6} \textbf{until convergence}
\end{algorithm}

\subsection{P2P Compression}
Next, we tackle \eqref{eq:problem-P2P} for the P2P compression. The access link rate constraint in \eqref{eq:problem-P2P-rate} can be readily addressed by the WMMSE formulation \eqref{eq:inequality-WMMSE-1}-\eqref{eq:optimal-w-u}. Also, similar to the WZ compression case, it can be shown that \eqref{eq:fronthaul-constraint-P2P} is fulfilled if there exists a positive definite matrix $\boldsymbol{\Sigma}_i \in \mathbb{C}^{N\times N}$ such that
\begin{align}    &\log_2\det\left(\boldsymbol{\Sigma}_i\right) + \frac{1}{\ln 2} \text{tr}\left( \boldsymbol{\Sigma}_i^{-1} \left( \boldsymbol{\Omega}_i + \mathbf{D}_i^H \mathbf{V}_{\Sigma} \mathbf{D}_i \right) \right) - \frac{N}{\ln 2} \nonumber \\&\,\,\,\,\,\,  - \log_2\det\left(\boldsymbol{\Omega}_i\right) \leq \frac{rC_F}{L}, \, i\in\mathcal{L}. \label{eq:convexified-fronthaul-constraint-P2P}
\end{align}
Constraint (\ref{eq:convexified-fronthaul-constraint-P2P}) boils down to (\ref{eq:problem-P2P-fronthaul}) with the auxiliary matrix $\boldsymbol{\Sigma}_i$ set to
\begin{align}
    \boldsymbol{\Sigma}_i = \boldsymbol{\Omega}_i + \mathbf{D}_i^H \mathbf{V}_{\Sigma} \mathbf{D}_i. \label{eq:optimal-Sigma-P2P}
\end{align}
As a consequence, an equivalent formulation of \eqref{eq:problem-P2P} for the P2P compression scheme is given as
\begin{align}
    \underset{\mathbf{v}, \boldsymbol{\Omega}, \mathbf{R}, \mathbf{u}, \mathbf{w}, \boldsymbol{\Sigma}} {\mathrm{max.}}\,\,\, & \sum\nolimits_{k\in\mathcal{K}} R_k \, \label{eq:problem-P2P-modified} \\
 \mathrm{s.t. }\,\,\,\,\,\, & R_k \ln 2 \leq \ln w_k - w_k \, e_k( \mathbf{v}, \boldsymbol{\Omega}, u_k ) + 1, \, k\in\mathcal{K}, \nonumber\\
 & \ln\det\left(\boldsymbol{\Sigma}_i\right) + \text{tr}\left( \boldsymbol{\Sigma}_i^{-1} \left( \boldsymbol{\Omega}_i + \mathbf{D}_i^H \mathbf{V}_{\Sigma} \mathbf{D}_i \right) \right) - N \nonumber\\
 & \,\,\,\,\,\,\,\,\,\,\,\,\,- \ln\det\left(\boldsymbol{\Omega}_i\right) \leq \frac{rC_F}{L} \ln 2,\,  i\in\mathcal{L}, \nonumber \\
 & \sum\nolimits_{k\in\mathcal{K}} \left\|\mathbf{D}_i^H\mathbf{v}_k\right\|^2 + \text{tr}\left(\boldsymbol{\Omega}_i\right) \leq P_{\text{tx}}, \, i\in\mathcal{L}, \nonumber 
\end{align}
where $\mathbf{\Sigma}=\{\mathbf{\Sigma}_{i}\}_{i\in\mathcal{L}}$. A stationary point of problem (\ref{eq:problem-P2P-modified}) can be efficiently found through an alternating optimization procedure between two blocks of variables $\{\mathbf{v},\boldsymbol{\Omega}\}$ and $\{\mathbf{u},\mathbf{w},\boldsymbol{\Sigma}\}$ as detailed in Algorithm 2.

\begin{algorithm}
\caption{Alternating optimization algorithm for problem (\ref{eq:problem-P2P-modified}) under the P2P compression}

\textbf{\footnotesize{}1}\textbf{ Initialize:} Set $t\leftarrow1$, and initialize $\{\mathbf{v},\boldsymbol{\Omega}\}$ to an arbitrary point that satisfies (\ref{eq:problem-P2P-rate})-(\ref{eq:problem-P2P-power});

\textbf{\footnotesize{}2}\textbf{ repeat}

\textbf{\footnotesize{}3}~~~~$t\leftarrow t+1$;

\textbf{\footnotesize{}4}~~~~update $\{\mathbf{u},\mathbf{w},\boldsymbol{\Sigma}\}$ according to (\ref{eq:optimal-w-u}) and (\ref{eq:optimal-Sigma-P2P});

\textbf{\footnotesize{}5}~~~update $\{\mathbf{v},\boldsymbol{\Omega}\}$
as a solution of the problem (\ref{eq:problem-P2P-modified}) with $\{\mathbf{u},\mathbf{w},\boldsymbol{\Sigma}\}$ being fixed;

\textbf{\footnotesize{}6} \textbf{until convergence}
\end{algorithm}

\section{Numerical Results} \label{sec:numerical}

This section presents numerical results assessing the proposed algorithms.
The UEs are randomly located within a circular area of radius 200 m, while the APs are positioned evenly along the boundary circle, maintaining equal angular intervals. For a given distance $d_{k,i}$ between UE $k$ and AP $i$, the path-loss $\rho_{k,i}$ of wireless access link between UE $k$ and AP $i$ is given as $\rho_{k,i} = \rho_0 ( d_{k,i}/d_0 )^{-\alpha}$ where $d_{k,i}$ accounts for the distance of the associated link, $\rho_0 = 0$ dB is the reference path-loss at $d_0 = 30$ m, and $\alpha = 3$ indicates the path-loss exponent.
Unless stated otherwise, the CSI error power $\beta_{k,i}$ is assumed to be $0.1 \rho_{k,i}$ which corresponds to the case where the error variance equals 10 $\%$ of the actual channel power.
We also assume that the fronthaul and access transmission phases span an equal time duration, i.e., $n_{A}=n_{F}$ and $r=1$.

\begin{figure}
\centering\includegraphics[width=0.8\linewidth]{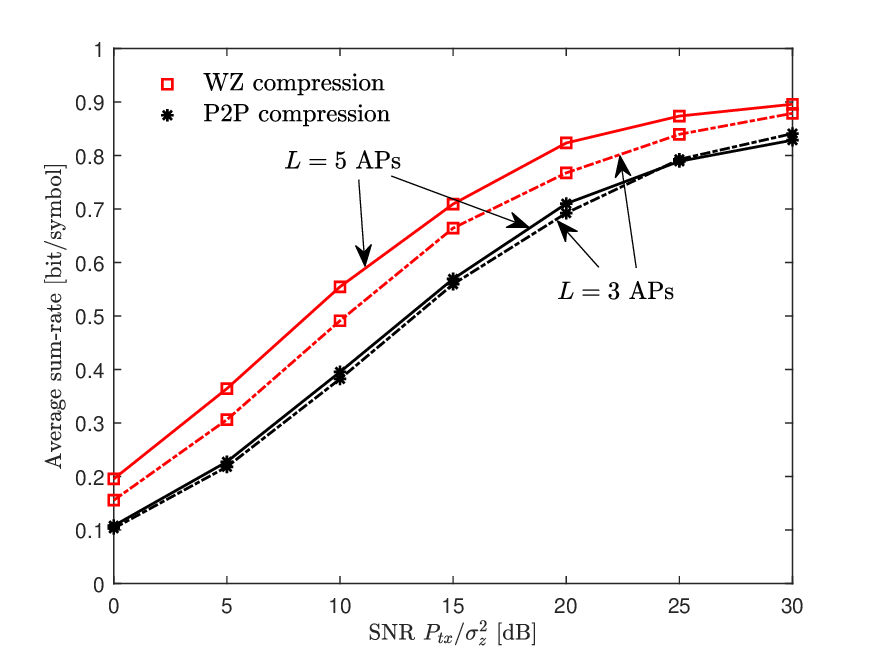}
\vspace{-5mm}
\caption{\label{fig:avgRsum-vs-SNR}Average sum-rate versus the SNR $P_{\text{tx}}/\sigma_z^2$ ($K=10$ UEs, $L\in\{3,5\}$ APs each with $N=2$ antennas, and $C_F=1$ bit/symbol)}
\end{figure}

\begin{figure}
\centering\includegraphics[width=0.8\linewidth]{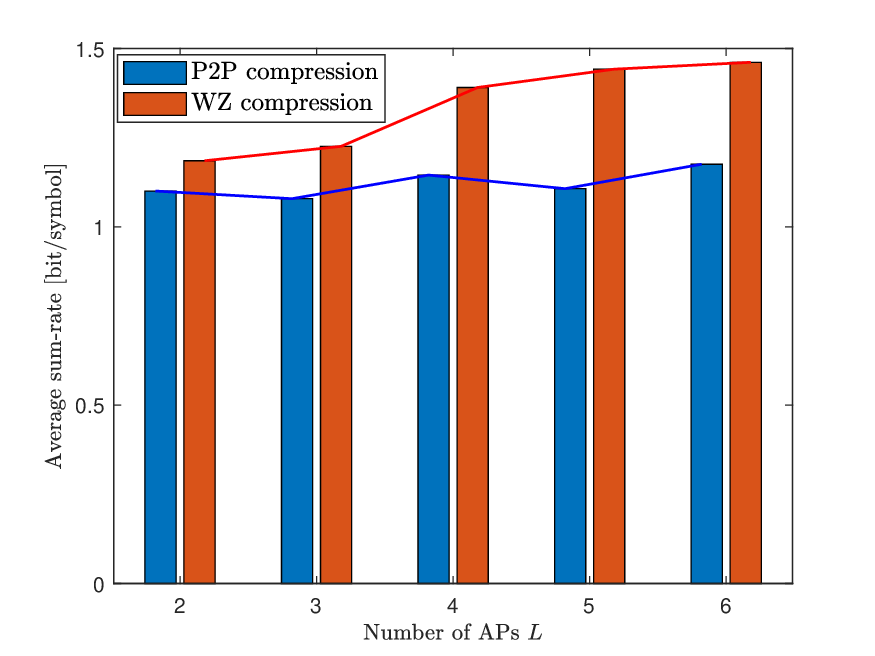}
\vspace{-5mm}
\caption{\label{fig:avgRsum-vs-L}Average sum-rate versus the number of APs $L$ ($K=10$ UEs, $N=2$ AP antennas, $C_F=2$ bit/symbol, and $P_{\text{tx}}/\sigma_z^2=15$ dB)}
\end{figure}

In Fig. \ref{fig:avgRsum-vs-SNR}, we plot the average sum-rate with respect to the transmit SNR $P_{\text{tx}} / \sigma_z^2$ with $K=10$ UEs, $L\in\{3,5\}$ APs each with $N=2$ antennas, and $C_F=1$ bit/symbol.
The sum-rate performance of both compression schemes increases with the SNR while saturating to finite levels due the finite-capacity fronthaul links.
Also, the WZ compression scheme, in which the APs utilize the side information for decompression, provides notable gains compared to the P2P compression scheme.
It should be emphasized that increasing the number $L$ of APs has conflicting impacts on the sum-rate. This is because, although
deploying more APs brings a beamforming gain, the effective compression rate available for each sample will decrease (see the right-hand sides of (\ref{eq:fronthaul-constraint-P2P}) and (\ref{eq:fronthaul-constraint-WZ})), since the fronthaul phase is divided into $L$ time slots, where each slot is used for delivering a single fronthaul block.
For this reason, the sum-rate of the P2P compression remains almost same when the number of APs increases from $L=3$ to $5$.
Instead, the performance of the WZ scheme is significantly improved with more APs, because the degraded quantization fidelity is overcome by utilizing the side information for decompression.
We validate this pattern once again in Fig. \ref{fig:avgRsum-vs-L} by plotting the average sum-rate with respect to the number of APs $L$ for a cell-free MIMO system with $K=10$ UEs, $N=2$ AP antennas, $C_F=2$ bit/symbol, and $P_{\text{tx}}/\sigma_z^2=15$ dB.
Similar to Fig. \ref{fig:avgRsum-vs-SNR}, we observe that as the number of APs $L$ increases, only the sum-rate of the WZ scheme shows improvement.

\begin{figure}
\centering\includegraphics[width=0.8\linewidth]{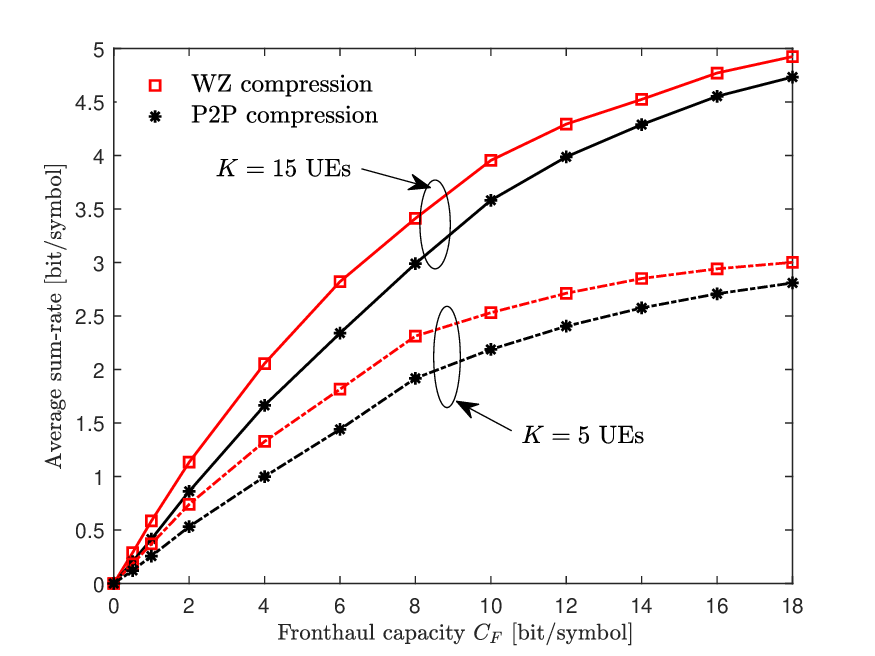}
\vspace{-5mm}
\caption{\label{fig:avgRsum-vs-CF}Average sum-rate versus the fronthaul capacity $C_F$ ($K\in\{5,15\}$ UEs, $L=4$ APs each with $N=2$ antennas, and $P_{\text{tx}}/\sigma_z^2=10$ dB)}
\end{figure}

Fig. \ref{fig:avgRsum-vs-CF} shows the impact of the fronthaul capacity $C_F$ on the average sum-rate for the downlink of a cell-free MIMO system with $K\in\{5,15\}$ UEs, $L=4$ APs each with $N=2$ antennas, and $P_{\text{tx}}/\sigma_z^2=10$ dB.
The sum-rates of both the P2P and WZ schemes grow with the fronthaul capacity $C_F$, since with a larger $C_F$, the fronthaul quantization causes a smaller distortion.
The performance gain of the WZ scheme is most pronounced in the intermediate regime of $C_F$, which means that the advantage of the advanced WZ compression is minor when $C_F$ is sufficiently large or small.

\begin{figure}
\centering\includegraphics[width=0.8\linewidth]{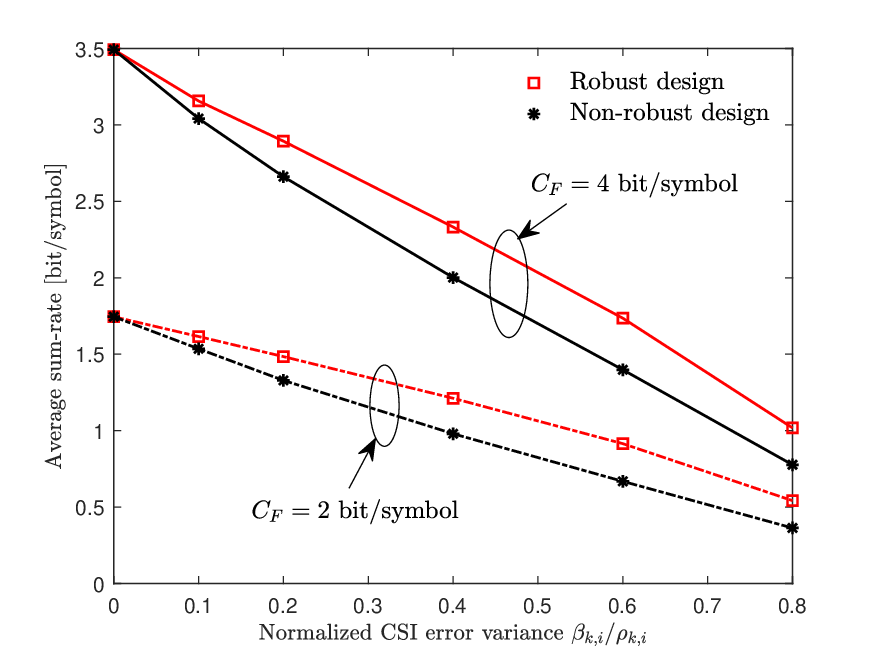}
\vspace{-5mm}
\caption{\label{fig:avgRsum-vs-beta}Average sum-rate versus the normalized CSI error variance $\beta_{k,i}/\rho_{k,i}$ ($K=10$ UEs, $L=4$ APs each with $N=2$ antennas, $C_F\in\{2,4\}$ bit/symbol and $P_{\text{tx}}/\sigma_z^2=20$ dB)}
\end{figure}

Fig. \ref{fig:avgRsum-vs-beta} assesses the significance of the proposed robust design, which incorporates the CSI error in the optimization process, by comparing the average sum-rate of the proposed WZ scheme with that of a non-robust design. The non-robust design optimizes the precoding vector $\mathbf{v}$ and the fronthaul compression strategy $\boldsymbol{\Omega}$ without considering CSI error, but its average sum-rate is evaluated using the true CSI error variance.
We plot the average sum-rate performance of the proposed robust and the baseline non-robust schemes while gradually increasing the normalized CSI error variance, defined as $\beta_{k,i}/\rho_{k,i}\in [0,1]$, for a cell-free MIMO system with $K=10$ UEs, $L=4$ APs each with $N=2$ antennas, $C_F\in\{2,4\}$ bit/symbol and $P_{\text{tx}}/\sigma_z^2=20$ dB.
With the increased CSI errors, the performance of the non-robust scheme experiences a significant degradation. Nevertheless, by incorporating CSI imperfections into the design of precoding and fronthaul compression strategies, it is feasible to mitigate the performance degradation.
We also observe that the performance gap between the robust and non-robust schemes becomes more evident as $C_F$ increases.

\section{Conclusion} \label{sec:conclusion}

We have studied the downlink of a cell-free MIMO system with radio stripes fronthaul network.
Noting the serial transfer of fronthaul-compressed block through APs, an application of WZ compression was discussed whereby each AP decompresses all the received blocks, that pass through it, and utilize them for decompressing its desired precoded signal.
For both the standard P2P and advanced WZ compression, we tackled the sum-rate maximization problems under per-AP power and fronthaul capacity constraints.
Via numerical results, the advantages of the WZ compression compared to the P2P compression were validated.
As future work, we consider an extension to the segmented fronthaul network in which multiple, not a single, radio stripes are connected in parallel \cite{Fernandes:WCL22}.
Furthermore, an intriguing and demanding subject would involve the development of a deep learning-based sequential linear processing of APs that ensures scalability with respect to the numbers of UEs and APs \cite{Lee:TWC21}.

\end{document}